\documentclass[a4paper,11pt]{article}
\pdfoutput=1
\usepackage{jheppub} 

\usepackage{graphicx}  
\usepackage{dcolumn}   
\usepackage{bm}        
\usepackage{amssymb}   
\usepackage{amsmath}
\usepackage{xspace}
\usepackage{hyperref}

\usepackage{lineno}
\newcommand{\pttop}{\ensuremath{p_\mathrm{T}^{\mathrm{top}}}\xspace}
\newcommand{\pttopre}{\ensuremath{p_\mathrm{T}^{\mathrm{top,reco}}}\xspace}
\newcommand{\mjet}{\ensuremath{m_\mathrm{jet}}\xspace}
\newcommand{\mtop}{\ensuremath{m_\mathrm{t}}\xspace}
\newcommand{\mtopre}{\ensuremath{m_\mathrm{t}^{\mathrm{reco}}}\xspace}

\newcommand{\MADGRAPH} {\textsc{MadGraph}\xspace}
\newcommand{\MCATNLO} {\textsc{mc@nlo}\xspace}
\newcommand{\MGvATNLO}{\MADGRAPH{}5\_a\MCATNLO}
\newcommand{\pt}{\ensuremath{p_{\mathrm{T}}}\xspace}
\newcommand{\ttbar}{\ensuremath{t\bar{t}}\xspace}
\newcommand{\xL}{\ensuremath{x_\mathrm{L}}\xspace}
\newcommand{\mw}{\ensuremath{m_{W}}\xspace}
\newcommand{\fbinv} {\mbox{\ensuremath{\,\text{fb}^{-1}}}\xspace}
\newcommand{\GeV}{\ensuremath{\,\text{Ge\hspace{-.08em}V}}\xspace}

\title{Prospect of measuring the top quark mass through energy correlators}

\author{Meng Xiao, }
\author{Yulei Ye, }
\author{and Xinyu Zhu}
\affiliation{School of Physics, Zhejiang University, Hangzhou, Zhejiang 310058, China}

\emailAdd{mxiao@zju.edu.cn}
\emailAdd{ylye@zju.edu.cn}
\emailAdd{xy\_zhu@zju.edu.cn}

\abstract{
Reaching a high precision of the top quark mass is an important task of the Large Hadron Collider. We perform a feasibility study of measuring the top quark mass through the three-point energy correlator. The expected sensitivity of the top quark mass in the boosted regime is presented. We further introduce its application to the low top \pt regime and demonstrate that both the W boson and the top quark masses could be extracted from this single observable. Compared to traditional observables, the energy correlator shows robustness to uncertainties that usually dominate experimental measurements and provides a promising way to improve experimental precision.
}

\begin{document}

\maketitle 
\flushbottom

\section{\label{sec:intros}Introduction}

The top quark is the heaviest elementary particle in the Standard Model. The precision of its mass, together with that of the Higgs boson, plays a central role in determining the stability of the electroweak vacuum~\cite{Alekhin:2012py, Degrassi:2012ry, Buttazzo:2013uya, Andreassen:2017rzq}.
While the precise measurement of the top quark mass has been one of the most important campaigns at the Large Hadron Collider (LHC), the yielded uncertainty is the largest among those of the heavy elementary particles~\cite{ParticleDataGroup:2022pth}.

Various strategies have been developed to precisely measure the top quark mass \mtop \cite{CMS:2024irj, ATLAS:2018fwq, CMS:2015lbj, CMS:2018tye, CMS:2023ebf, CMS:2019fak, CMSk2021jnp, CMS:2022emx, ATLAS:2017lox, CMS:2016iru, CMS:2016ixg, ATLAS:2014nxi, ATLAS:2022jbw, CMS:2022kqg}, among which the most precise one comes from the direct measurements \cite{CMS:2023ebf}, where the top quark is treated as a free particle and Monte-Carlo (MC) based simulation is used to extract the \mtop. 
The experimental signature and the corresponding measurement vary with the transverse momentum of the top quark \pttop. The low \pttop region, often called the resolved regime, presents a final state with three particles decayed from the top quark and well-separated from each other.
Experiments usually reconstructed the invariant masses from the three objects to measure the \mtop.
As the \pttop increases, the decayed products get Lorentz-boosted, making the hadronic decay products of the top quark non-resolvable and forming a single large jet. Typical observables for the \mtop measurement in this regime are the invariant jet mass. 

For both \pttop regimes, the reconstructed invariant masses are subject to uncertainties in the jet $p_T$, such as the jet energy scale uncertainty (JES).
 Although the JES has been constrained experimentally by the precisely known W boson mass~\cite{CMS:2023ebf, CMS:2022kqg}, it still dominates the uncertainties in these measurements.
Recently, energy correlators inside jets have been proposed to study the properties of heavy particles such as the top quark mass\cite{Holguin:2022epo, Holguin:2023bjf} in the boosted regime. They have the advantage of being theoretically calculable and, therefore could be used to extract the pole mass of the top quark. These correlators have been measured for light flavor jets by the CMS collaboration \cite{CMS:2024mlf}, demonstrating that high-precision measurements of the observable could be reached.

In this paper, we illustrate that the energy correlator is an ideal observable to reduce the impact of JES in top mass measurements. It is powerful to improve the precision of the \mtop not only in the boosted regime but also in the resolved regime. 
We first present its sensitivity to the \mtop in the boosted region using MC simulations, where the projected three-point energy correlator (E3C) is compared with the jet invariant mass \mjet, and better resilience to systematic uncertainties such as JES is observed.
We then extend the method to the low \pttop regions, where two peaks resulting from the W boson and the top quark could be observed simultaneously in the E3C distribution. The intriguing features of the peaks make the E3C a promising observable to improve the \mtop precision in the resolved region.

\section{\label{sec:eec}E3C and its application to top quark}

Multi-point energy correlators \cite{Basham:1978bw, Dixon:2019uzg, Lee:2022ige} are a class of observables that describe the energy-weighted angular correlations between particles.
Specifically, the E3C focuses on the correlation among three particles. The theoretical definition of E3C is \cite{Chen:2020vvp, Chen:2023zlx}

\begin{equation}
    \text{E3C} = \frac{d\sigma}{dx_{\text{L}}}
    =  \sum_{i,j,k}^{n} \int d\sigma
    \frac{E_{i}E_{j}E_{k}}{E^3} 
     \delta \left( x_{\text{L}} - \text{max}
        \left( \Delta R_{ij}, \Delta R_{ik}, \Delta R_{jk}  \right)
    \right).
    \label{eq:E3C}
\end{equation}

At hadron colliders, they are proposed for jet substructure studies. In its application to boosted top jets \cite{Holguin:2022epo}, the particle indices $i,j,k$ run overall all the $n$ particles in a top jet, and $E$ is the summed energy of all the particles, equivalent to the jet energy.
The largest $\Delta R_{ij} = \sqrt{\Delta\eta^2+\Delta\phi^2}$ of the triangle formed by $i,j,k$ particles is denoted by \xL. Since the top quark decays to three particles, E3C is a natural probe for the top decay. The distribution of E3C is a function of \xL with energy weight $E_iE_jE_k/E^3$. 
Experimental resolution of angles is much better than the energies, therefore compared to invariant mass type observables, E3C could be measured with higher precision. In addition, the weight makes the observable robust to uncertainties that systematically change the energy of the jet constituents, since the changes largely cancel in the ratio.

\section{\label{sec:boost}Top mass sensitivity in the boosted region}
We start by examining the sensitivity of the E3C observable to the \mtop in the boosted top region. MC simulation of the $t\bar{t}$ semi-leptonic process is used. According to experimental measurements \cite{CMS:2024yqd}, this process yields the highest sensitivity.
The events are generated with \MGvATNLO~\cite{Alwall:2014hca, Frederix:2018nkq} at leading-order (LO) of QCD and interfaced to PYTHIA8 \cite{Bierlich:2022pfr} for parton shower and hadronization. Multiple samples are generated with different top quark masses, ranging from 168 \GeV to 174 \GeV. The mass of the W boson is set to 80.4 \GeV. To select the hadronically decayed top quark, anti-$k_{\mathrm{T}}$ algorithm \cite{Cacciari:2008gp} is performed by FastJet~\cite{Cacciari:2005hq, Cacciari:2011ma} with parameter $R=1.2$. An event is required to have at least two jets with $\pttop > 400$ GeV and $\lvert \eta \rvert < 2.4$ and a lepton with $\pt > 60 \GeV$ and $\lvert \eta \rvert < 2.4$. The leading jet is required to be further away from the lepton compared to the sub-leading jet, and it is used to evaluate the sensitivity to the \mtop.

Here we compare the performances of E3C to \mjet. CMS has used \mjet to measure the \mtop with a relative uncertainty of 1.45\%~\cite{CMS:2019fak} with 36\fbinv data and the uncertainty was further reduced to 0.50\%~\cite{CMS:2022kqg} after constraining JES uncertainty using the W mass with 138~\fbinv data. We have the choice to extract the \mtop either from the reconstructed shapes at the detector level or from the unfolded shapes at the generator level. The advantage of the latter, as proposed in Ref~\cite{CMS:2019fak}, is that once theoretical calculation is available, it could be used to extract the pole mass instead of the MC mass. For this reason, here we use the generator distribution to derive the sensitivity. To have a realistic estimation of selection efficiency and statistical loss due to the unfolding procedure, we downscale the number of events so that the statistical uncertainty of \mtop derived from \mjet is the same as the CMS result in Ref.~\cite{CMS:2019fak}. For the comparison between the two variables, we assume the unfolding impact is similar and the same downscale factor is applied to both distributions.

\begin{figure}[ht]
    \centering
        \includegraphics[width=0.48\linewidth]{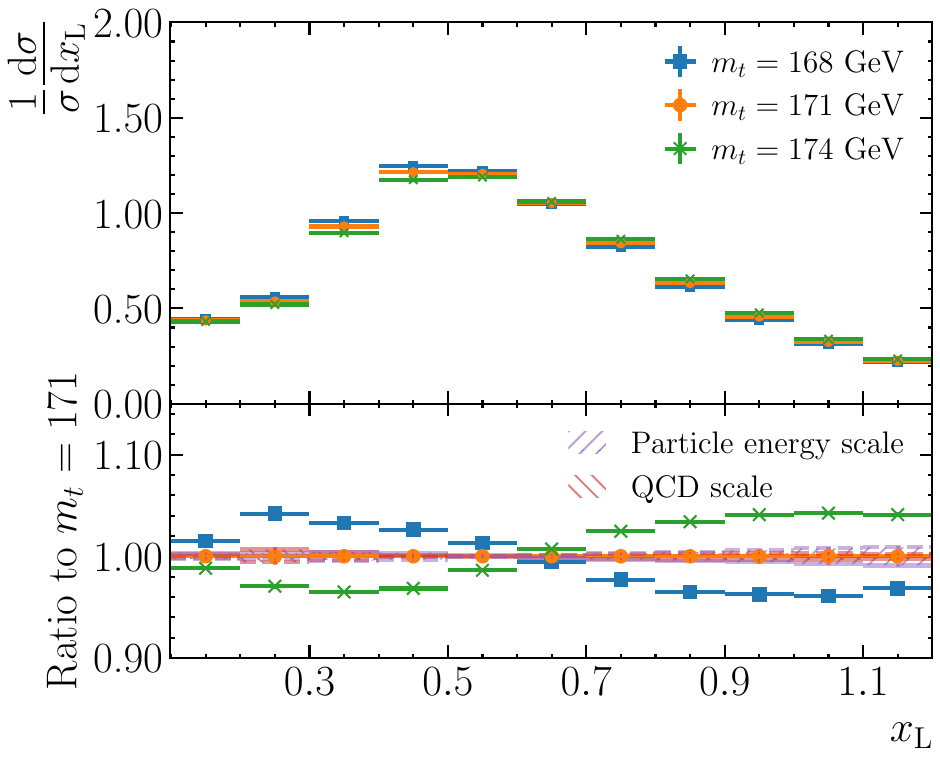} 
        \includegraphics[width=0.48\linewidth]{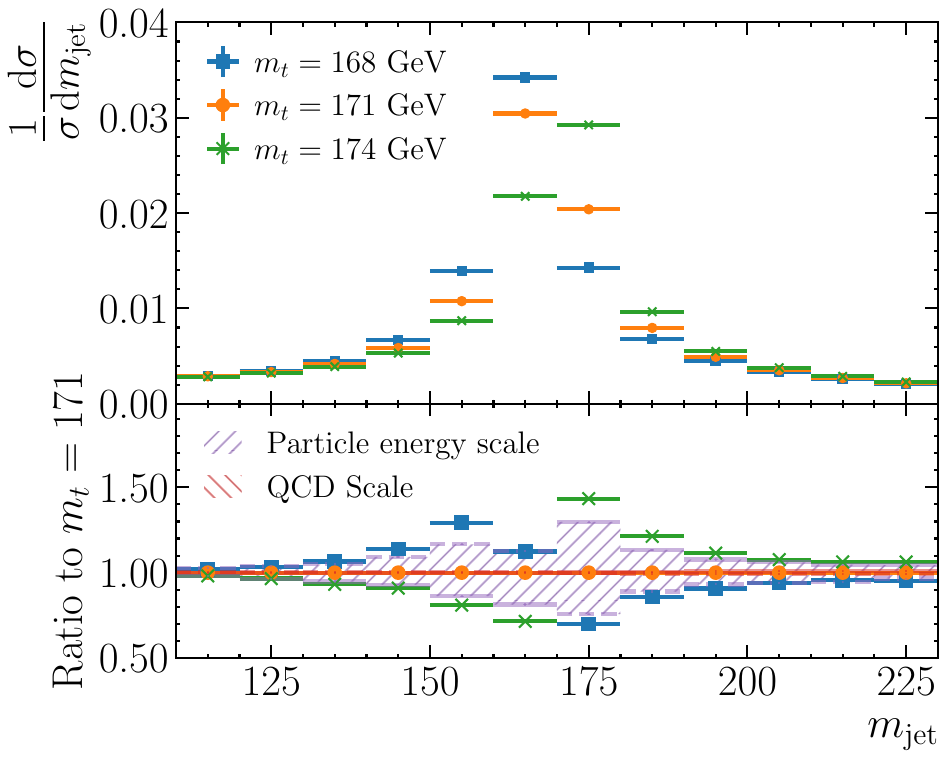}
    \caption{
    The distribution of E3C (left) and \mjet (right) of boosted top jets with different \mtop values and their ratio to $\mtop=171 \GeV$. The leading systematic uncertainties are shown in hatched bands in the ratio panel. Here the photon, charged, and neutral particle energy scales are combined into the total particle energy scale. 
    }
    \label{fig:boosted_e3c}
\end{figure}

Figure~\ref{fig:boosted_e3c} shows the E3C and \mjet distribution built from the boosted jet. All the hadrons inside the jet are used for the calculation. A peak is observed for both observables, which originated from the top quark decay, and is sensitive to the \mtop, as shown by the distributions of various \mtop values. For the \mjet, this is rather straightforward to understand. As for the E3C, it could be explained by $\xL \propto \mtop/\pttop$~\cite{Holguin:2022epo} in large \pttop limit.
Since \pttop remains relatively constant as $m_t$ changes, the peak position is directly proportional to the \mtop.
Compared to E3C, the nominal shape of \mjet is more sensitive to the \mtop. The variation caused by a 3 \GeV change in \mtop is about 40\% compared to 5\% in the E3C. However, the same variation in \mjet could be caused by the jet \pt scale, as shown by the shaded distribution in figure~\ref{fig:boosted_e3c}. This is the reason why JES uncertainty plays an important role in such measurements~\cite{CMS:2019fak}. On the other hand, the figure shows that E3C is much less affected by this uncertainty.

To evaluate the impact of various systematic uncertainties, we consider the sources that potentially dominate the two measurements from Ref.~\cite{CMS:2019fak} and ~\cite{CMS:2024mlf}. This includes the energy scale uncertainties of the particles within the jet, which is 1\% for the charged particles, 3\% for the photons and 5\% for the neutral particles~\cite{CMS:2024mlf}.
Each source affects the particles of a particular type and, consequently the overall jet \pt. The combined particle energy scale effect is analogous to the JES uncertainty, which changes the jet \pt by approximately 1-2\% for the jets considered here.
The initial-state (ISR) and final-state radiation (FSR) uncertainties are considered by varying the renormalization scale $\mu$ in parton shower by 1/2 and 2.
The QCD scale uncertainties that take into account the missing higher-order calculation in hard scattering processes are obtained from varying the renormalization and factorization scales independently by a factor of 0.5 and 2. 

\begin{figure}[t!]
    \centering
    \includegraphics[width=0.8\linewidth]{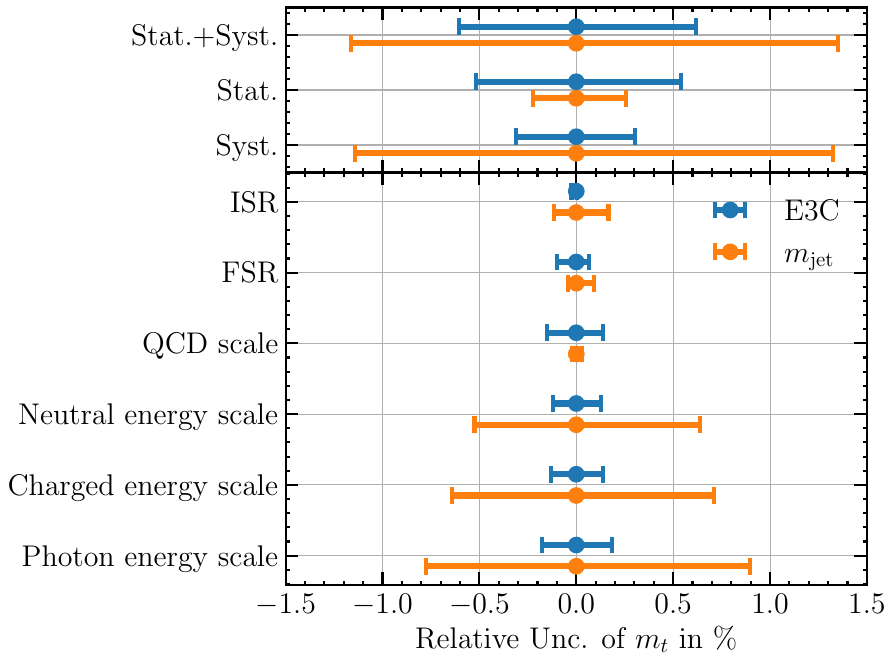}
    \caption{
    The expected uncertainties of \mtop (in \% of $\mtop=171$ GeV) using E3C and \mjet distributions, at $\mathcal{L}=36 \fbinv$. The statistical uncertainties and a breakdown of the systematic uncertainties are shown.
    }
    \label{fig:boosted_uncert}
\end{figure}

We calculate the $\chi^2$ values as a function of the \mtop under the impact of the above uncertainties for \mjet and E3C respectively. We take the shape of $\mtop=171\GeV$ as the nominal, and a deviation of $\Delta \chi^2=1$ is used to extract the uncertainties. An integrated luminosity of 36~\fbinv is used. The estimated uncertainties from this analysis are presented in figure ~\ref{fig:boosted_uncert}. The \mjet result is dominated by systematic uncertainties of the particle energy scales inside the jet, while the E3C based result is mainly limited by the statistical uncertainty. At the luminosity of 36 \fbinv, the E3C already yields a better overall sensitivity of 0.6\% compared to 1.2\% from \mjet. It is expected that the E3C based result would gain more from increased statistics. At the luminosity of 300 \fbinv, the statistical uncertainty for E3C will decrease to 0.2\%, comparable to the systematic uncertainty. 

\section{E3C in the resolved regime}
The most precise measurement of the \mtop was derived in the low \pttop region~\cite{CMS:2023ebf}, benefiting from the large cross section. The measurement was systematic dominated and the JES was one of the major sources. The above studies in the boost regime have demonstrated that the E3C observable is much less affected by this uncertainty. The overall \mtop precision would benefit more if the method could be extended to the low \pttop region. Here the hadronically decayed top quark is no longer contained in a single jet, but rather becomes three well separated jets. Therefore we modify the E3C definition and treat the three jets as the constituents of the top quark, and the $E$ in Eq.~\ref{eq:E3C}, which was the energy of the top quark jet, now becomes the sum of the energy of the three jets. Given that the four-momentum of a quark is well represented by the corresponding jet, this definition is even less affected by non-perturbative effects. As a start, we use the kinematic information of the three quarks decayed from the top quark to demonstrate the proof of principle.

We use the semi-leptonic \ttbar process to check the feasibility. Events are simulated with \MGvATNLO at LO of QCD. We focus on the events with $\pttop < 400 \GeV$. The E3C is built from the three quarks that hadronically decayed from the top quark. As the \pttop becomes comparable with the top quark mass, the dependence of the \xL on \pttop and \mtop becomes nontrivial. To differentiate the impact of the \pttop and \mtop, we split the events into slices of \pttop.

\begin{figure}[htbp]
     \centering
     \includegraphics[width=0.72\linewidth]{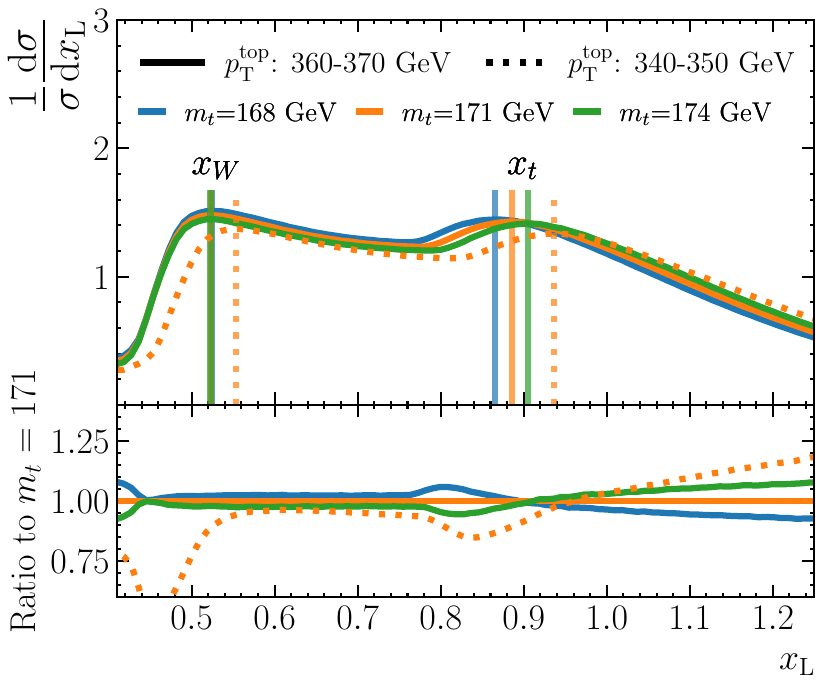}
     \caption{
     E3C distributions in the resolved regime in the range of $360 < \pttop < 370 \GeV$ and $340 < \pttop < 350 \GeV$. The three quarks decayed from the top quark are used to build E3C. A comparison between the distributions with different \mtop values is presented.
     }
     \label{fig:resolved_e3c}
\end{figure}

Figure~\ref{fig:resolved_e3c} shows the E3C distribution in two \pttop ranges: 360--370 \GeV and 340--350 \GeV. In each \pttop range, two distinct peaks are observed, and a structure arises from distinct decays of the W boson and top quark. We simulate events with different \mtop values while keeping the \mw fixed to 80.4 \GeV, as shown in figure~\ref{fig:resolved_e3c}. The right peak shifts with the \mtop while the left peak stays unchanged. This feature makes the E3C a nice observable to calibrate the \mtop through \mw. In earlier experimental measurements, this was achieved by constructing several observables from different inputs~\cite{CMS:2023ebf}. However, the dominant systematic uncertainties in extracting \mtop and \mw may not fully overlap and are well constrained in multiple observables. While in the case of E3C, consistent inputs are used to construct a single observable, resulting in a better constraint on the uncertainties. 

Similar to the approach in the boosted regime, the shape of the E3C can be used to measure \mtop in the resolved regime. The overall E3C distribution depends on \pttop as shown in figure~\ref{fig:resolved_e3c}, therefore a joint measurement of E3C across multiple \pttop regions could improve the sensitivity. However, this also makes the measurement potentially prone to the JES: a genuine $\pttop=345\GeV$ event, corresponding to the dashed curve in figure~\ref{fig:resolved_e3c}, might be reconstructed as $\pttop=365\GeV$ due to higher JES. 
If the distribution exhibited a single peak, similar to the often-used reconstructed \mtop in experimental measurements, this curve could be interpreted as either a higher \mtop or a higher JES, causing a large systematic uncertainty on the \mtop. However, the W peak on the left helps to disentangle the two effects: both peaks move with JES, and only the top peak shifts with \mtop variation, as shown in the ratio panel of figure~\ref{fig:resolved_e3c}. The two peak structure of E3C helps to reduce the impact of JES uncertainty significantly and provide a new method for experimental measurements. 

\begin{figure}[htbp]
     \centering
     \includegraphics[width=0.8\linewidth]{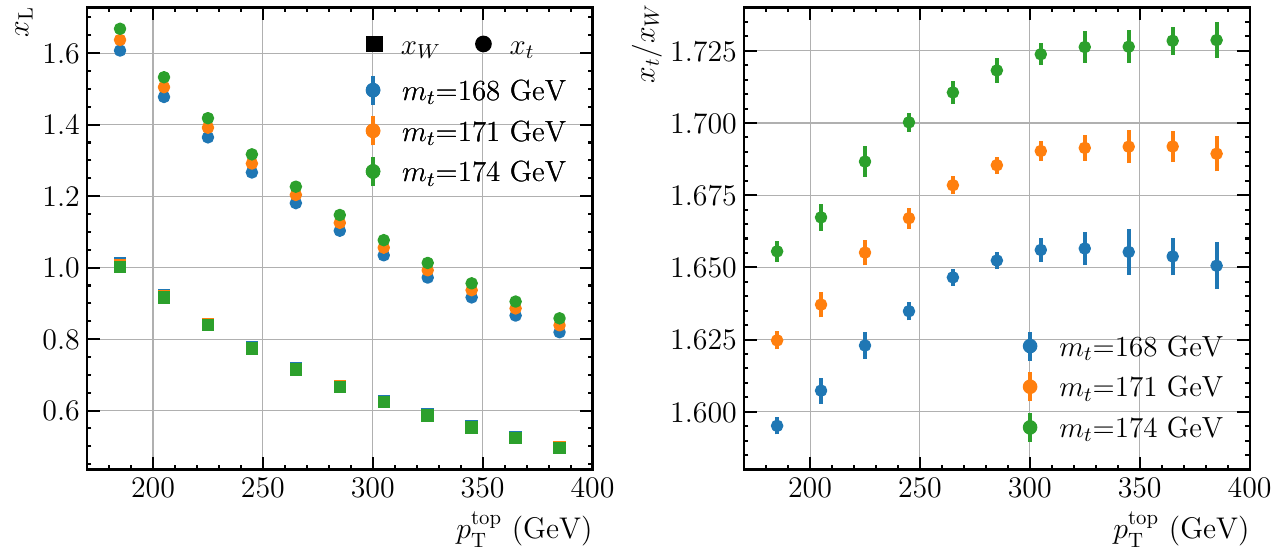}
     \caption{
     The peak position of $x_W$ and $x_t$ (left) and $x_t / x_W$ (right) in the E3C distributions as a function of average \pttop. The values are fitted from distributions under different \mtop assumptions.
     }
     \label{fig:resolved_peaks}
\end{figure}

The observation that both peaks in the E3C distribution shift together with \pttop raises the question of whether the ratio between them is less sensitive to \pttop. This could be explored to further reduce the uncertainty. To quantify that, we perform a fit to identify the \xL value corresponding to the two peaks, which we denote as $x_W$ and $x_t$ respectively.
We use an exponentially modified Gaussian function in LMFIT package~\cite{newville_2015_11813} to fit the E3C distribution near the peaks. The fit range of \xL is determined by requiring the $\chi^2 / \text{n. d.f.} < 2$. To take into account the impact of the fit range, we vary the nominal range by 10\% and take the difference between the determined peaks from the fits as a systematic uncertainty. The results are shown in figure \ref{fig:resolved_peaks}. The left figure shows the extracted $x_t$ and $x_W$ as a function of the average \pttop in each region. Both $x_t$ and $x_W$ decreases with increasing \pttop. Under different \mtop assumptions, the $x_W$ overlaps while the $x_t$ shows a sensitivity to the \mtop value. The right figure shows the ratio of $x_t/x_W$ as a function of \pttop. The ratio has a rather strong dependency on \pttop in the relatively low \pt region, while in the range above 300 \GeV, it becomes flatter. Experimental measurements could use the $x_t/x_W$ ratio in this region to reduce the sensitivity to any uncertainties that cause migrations in the reconstructed \pttop.

So far, we have focused on introducing the idea of using E3C and its feature to measure the top quark mass in the resolved regime in the ideal case.
Complications are expected when applying it to experimental measurements, including correctly picking jets originated from the top quark and possible degradation of E3C shape arising from jet reconstruction. For a realistic assessment of the effects, we use PYTHIA8 and Delphes~\cite{delphes} to simulate parton shower, hadronization and detector effects, where CMS settings are used. Jets are reconstructed by anti-$k_{\mathrm{T}}$ method with distance parameter of 0.4. For a jet with 100 \GeV \pt, the energy resolution is 10\%. Events are selected requiring one lepton with $\pt>60$ \GeV and four jets with $\pt > 30$ \GeV, among which two should be b-tagged. All the objects are required to have $|\eta| < 2.4$. The hadronic-decayed top quark is reconstructed from the two light jets and a b-tagged jet. Between the two b-tagged jets, the one that gives a \mtopre closer to 171 GeV is picked. The selection is rather crude and about 55\% of the events are correctly reconstructed from the top decay quarks. 

\begin{figure}[htbp]
    \centering
        \includegraphics[width=0.48\linewidth]{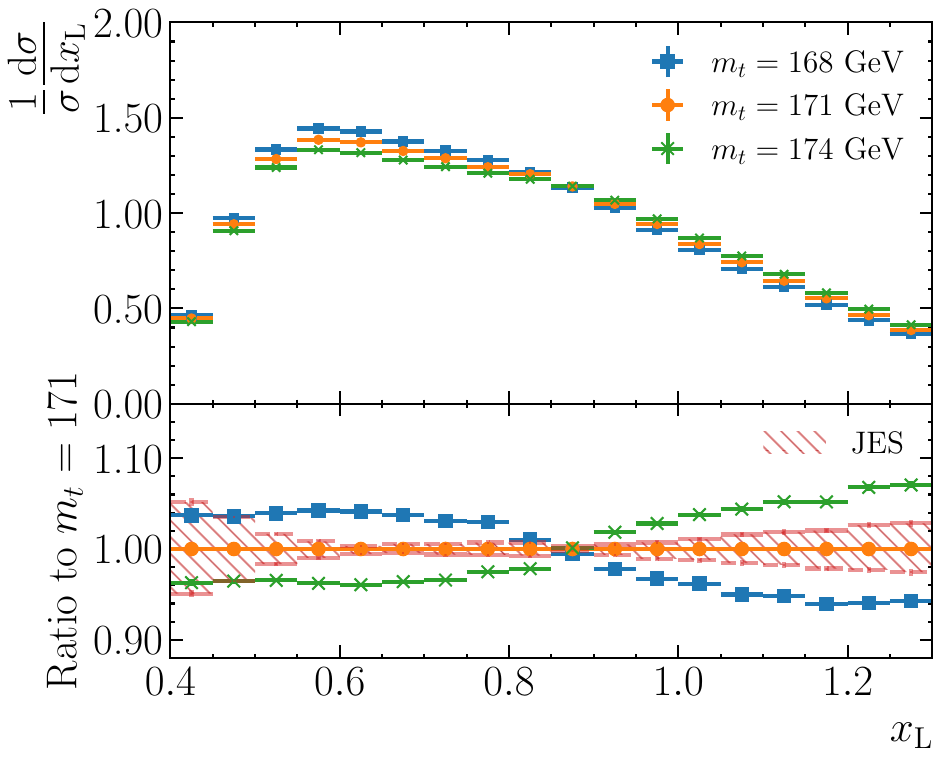} 
        \includegraphics[width=0.48\linewidth]{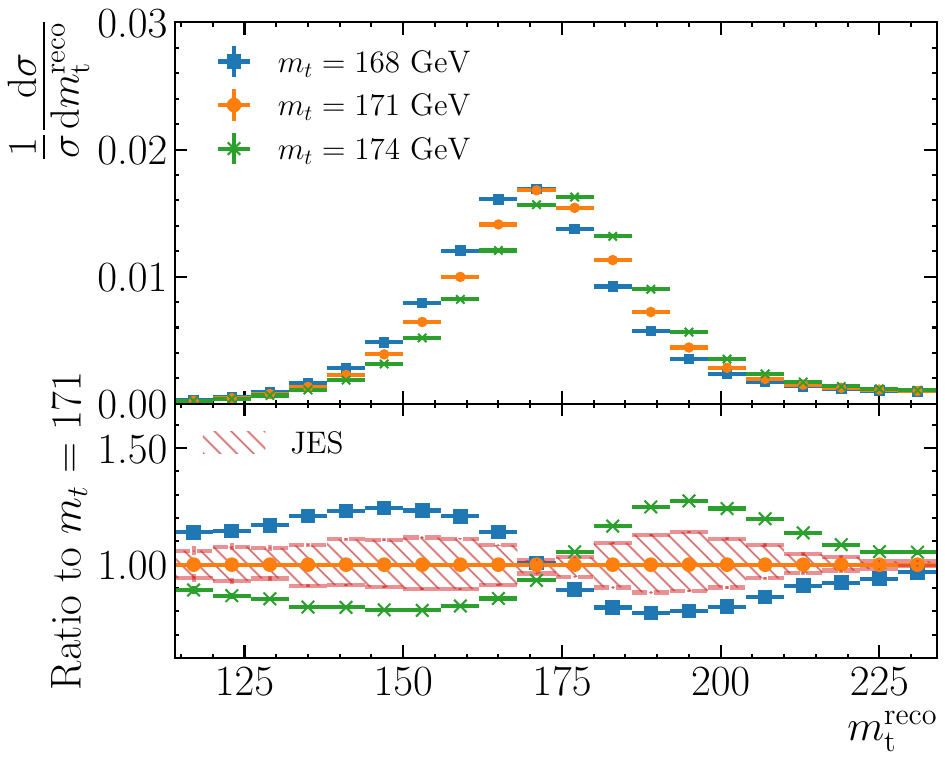}
    \caption{
    The normalized distribution of E3C (left) and \mtopre (right) reconstructed from three jets in the resolved region in the range of $360 < \pttopre < 370 \GeV$. Different \mtop values and their ratio to $\mtop=171 \GeV$ are shown. The JES uncertainty is shown in hatched bands in the ratio panel. 
    }
    \label{fig:resolved_reco}
\end{figure}

The E3C distribution built from the above selection is shown in figure~\ref{fig:resolved_reco}. The reconstructed $\pttop$ is required to be within 360--370 \GeV to compare with the ideal distribution in figure~\ref{fig:resolved_e3c}. It is clear that the two-peaks structure is smeared, however the shape dependency on \mtop remains. To evaluate the impact of JES, we take its uncertainty values from CMS~\cite{CMS-DP-2020-019}, which varies from 2--1\% for jets with \pt of 30 -- 100 \GeV, and 0.8\% for \pt $> 100$ \GeV. This causes event migrations between categories and a shape distortion. Figure~\ref{fig:resolved_reco} shows the E3C shape variation due to the JES, which is mild and different from that caused by \mtop. While for the $\mtop^{reco}$ the JES and \mtop effects are more similar.

\begin{figure}[htbp]
    \centering
        \includegraphics[width=0.48\linewidth]{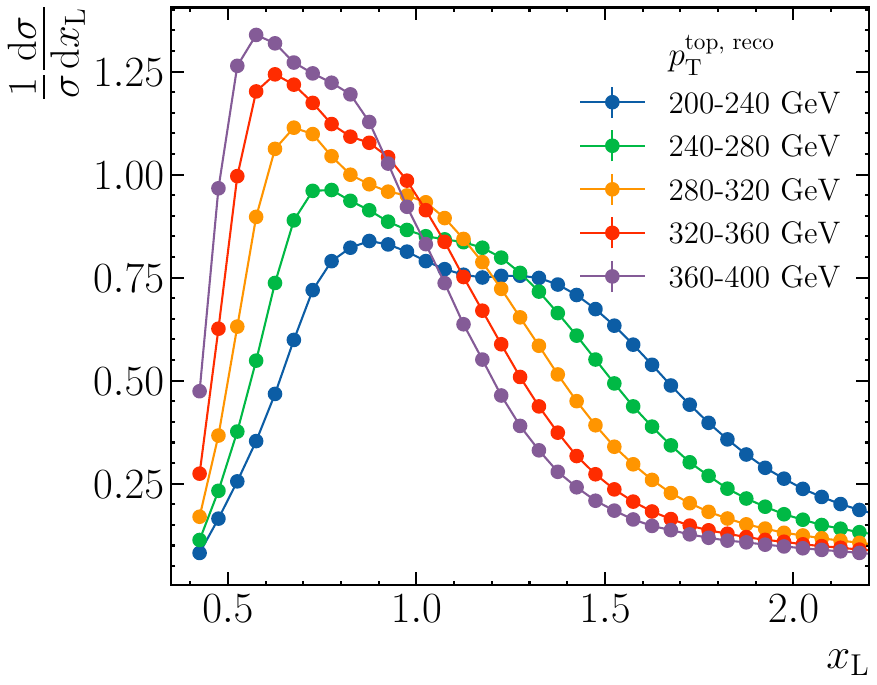} 
         \includegraphics[width=0.48\linewidth]{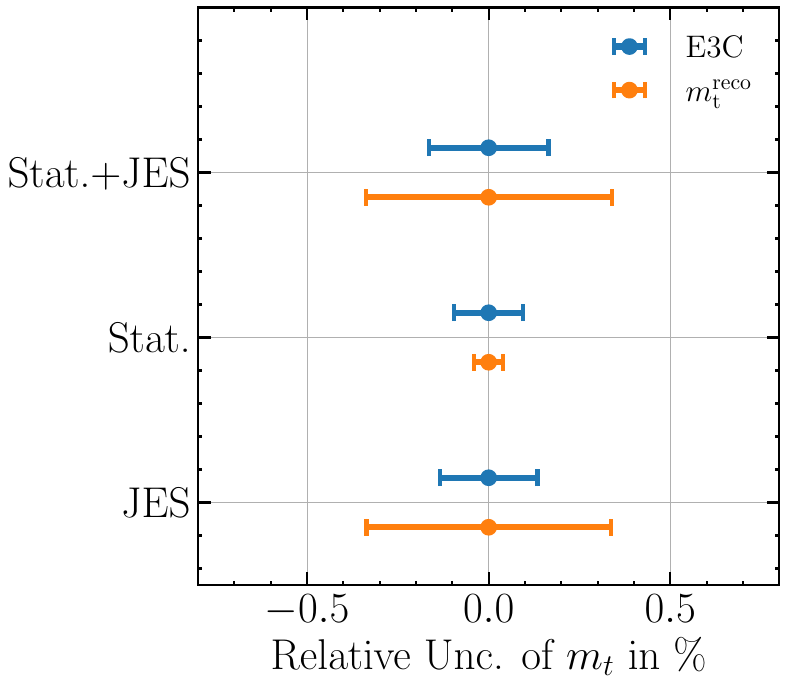}
    \caption{
   Left: the distribution of E3C reconstructed in five different \pttopre regions. Right: the expected uncertainties of \mtop (in \% of $\mtop=171$ GeV) using reconstructed E3C and \mtopre distributions in the resolved region, at $\mathcal{L}=137 \fbinv$. The statistical uncertainties and JES systematic uncertainties are shown.
    }
    \label{fig:resolved_5region}
\end{figure}

We use the reconstructed E3C distribution to check its sensitivity to \mtop under the JES impact, to be compared with $\mtop^{reco}$. Events with reconstructed \pttop in the range of 200--400 \GeV are selected and categorized in slices of \pttop, each covering a range of 40\GeV. The E3C distributions in the five categories are shown in figure~\ref{fig:resolved_5region}. Two peaks are visible in all the categories and shift with \pttopre. In principle finer bins could be adopted so that the peaks are sharper and help better constrain the JES. For simplicity here we use coarse binning. We calculate $\chi^2$ as a function of \mtop with JES as a nuisance parameter, which causes 2--7\% event migration in the considered categories and shape changes. The expected uncertainty of \mtop is shown in figure~\ref{fig:resolved_5region}. Similar to the conclusion in the boosted region, E3C is less sensitive to the JES uncertainty compared to the \mtopre. At the luminosity of 137 \fbinv, E3C already yields a better overall sensitivity. With such settings, a 0.1\% sensitivity is expected for E3C at 300 \fbinv. On top of this, improvements are expected to further increase the sensitivity, such as exploring lower \pttopre region, finer binning in categories and better top candidate selections. 

\section{Conclusions}

In summary, we explored the potential of using E3C to determine the top quark mass in both resolved and boosted regimes.
In the boosted regime, we showed that E3C provides a better mass sensitivity compared to the traditional jet mass observable in terms of systematic uncertainties, and is expected to benefit more from the increased statistics at the LHC.
In the resolved regime, the presence of two distinct peaks in the E3C distribution, corresponding to the top quark and the W boson, helps to calibrate the top quark mass and constrain jet energy scale uncertainties greatly.
These results demonstrate that E3C could be a promising observable for achieving high-precision measurement of the top quark mass.

\begin{acknowledgments}
We thank Huaxing Zhu for the useful discussions. The work is supported by National Natural Science Foundation of China (NSFC) under the Grant No. 12322504 and the center for high energy physics in Peking University.
\end{acknowledgments}

\bibliographystyle{JHEP}
\bibliography{top_mass_and_eec}

\providecommand{\noopsort}[1]{}\providecommand{\singleletter}[1]{#1}%

\providecommand{\href}[2]{#2}\begingroup\raggedright\begin{thebibliography}{10}

\bibitem{Alekhin:2012py}
S.~Alekhin, A.~Djouadi and S.~Moch, \emph{{The top quark and Higgs boson masses
  and the stability of the electroweak vacuum}},
  \href{https://doi.org/10.1016/j.physletb.2012.08.024}{\emph{Phys. Lett. B}
  {\bfseries 716} (2012) 214}
  [\href{https://arxiv.org/abs/1207.0980}{{\ttfamily 1207.0980}}].

\bibitem{Degrassi:2012ry}
G.~Degrassi, S.~Di~Vita, J.~Elias-Miro, J.R.~Espinosa, G.F.~Giudice, G.~Isidori
  et~al., \emph{{Higgs mass and vacuum stability in the Standard Model at
  NNLO}}, \href{https://doi.org/10.1007/JHEP08(2012)098}{\emph{JHEP} {\bfseries
  08} (2012) 098} [\href{https://arxiv.org/abs/1205.6497}{{\ttfamily
  1205.6497}}].

\bibitem{Buttazzo:2013uya}
D.~Buttazzo, G.~Degrassi, P.P.~Giardino, G.F.~Giudice, F.~Sala, A.~Salvio
  et~al., \emph{{Investigating the near-criticality of the Higgs boson}},
  \href{https://doi.org/10.1007/JHEP12(2013)089}{\emph{JHEP} {\bfseries 12}
  (2013) 089} [\href{https://arxiv.org/abs/1307.3536}{{\ttfamily 1307.3536}}].

\bibitem{Andreassen:2017rzq}
A.~Andreassen, W.~Frost and M.D.~Schwartz, \emph{{Scale Invariant Instantons
  and the Complete Lifetime of the Standard Model}},
  \href{https://doi.org/10.1103/PhysRevD.97.056006}{\emph{Phys. Rev. D}
  {\bfseries 97} (2018) 056006}
  [\href{https://arxiv.org/abs/1707.08124}{{\ttfamily 1707.08124}}].

\bibitem{ParticleDataGroup:2022pth}
{\scshape Particle Data Group} collaboration, \emph{{Review of Particle
  Physics}}, \href{https://doi.org/10.1093/ptep/ptac097}{\emph{PTEP} {\bfseries
  2022} (2022) 083C01}.

\bibitem{CMS:2024irj}
{\scshape CMS} collaboration, \emph{{Review of top quark mass measurements in
  CMS}},  \href{https://arxiv.org/abs/2403.01313}{{\ttfamily 2403.01313}}.

\bibitem{ATLAS:2018fwq}
{\scshape ATLAS} collaboration, \emph{{Measurement of the top quark mass in the
  $t\bar{t}\rightarrow $ lepton+jets channel from $\sqrt{s}=8$ TeV ATLAS data
  and combination with previous results}},
  \href{https://doi.org/10.1140/epjc/s10052-019-6757-9}{\emph{Eur. Phys. J. C}
  {\bfseries 79} (2019) 290}
  [\href{https://arxiv.org/abs/1810.01772}{{\ttfamily 1810.01772}}].

\bibitem{CMS:2015lbj}
{\scshape CMS} collaboration, \emph{{Measurement of the top quark mass using
  proton-proton data at ${\sqrt{(s)}}$ = 7 and 8 TeV}},
  \href{https://doi.org/10.1103/PhysRevD.93.072004}{\emph{Phys. Rev. D}
  {\bfseries 93} (2016) 072004}
  [\href{https://arxiv.org/abs/1509.04044}{{\ttfamily 1509.04044}}].

\bibitem{CMS:2018tye}
{\scshape CMS} collaboration, \emph{{Measurement of the top quark mass in the
  all-jets final state at $\sqrt{s} =$ 13 TeV and combination with the
  lepton+jets channel}},
  \href{https://doi.org/10.1140/epjc/s10052-019-6788-2}{\emph{Eur. Phys. J. C}
  {\bfseries 79} (2019) 313}
  [\href{https://arxiv.org/abs/1812.10534}{{\ttfamily 1812.10534}}].

\bibitem{CMS:2023ebf}
{\scshape CMS} collaboration, \emph{{Measurement of the top quark mass using a
  profile likelihood approach with the lepton~+~jets final states in
  proton\textendash{}proton collisions at $\sqrt{s}=13\,\text
  {Te}\hspace{-.08em}\text {V} $}},
  \href{https://doi.org/10.1140/epjc/s10052-023-12050-4}{\emph{Eur. Phys. J. C}
  {\bfseries 83} (2023) 963}
  [\href{https://arxiv.org/abs/2302.01967}{{\ttfamily 2302.01967}}].

\bibitem{CMS:2019fak}
{\scshape CMS} collaboration, \emph{{Measurement of the Jet Mass Distribution
  and Top Quark Mass in Hadronic Decays of Boosted Top Quarks in $pp$
  Collisions at $\sqrt{s} =$ 13 TeV}},
  \href{https://doi.org/10.1103/PhysRevLett.124.202001}{\emph{Phys. Rev. Lett.}
  {\bfseries 124} (2020) 202001}
  [\href{https://arxiv.org/abs/1911.03800}{{\ttfamily 1911.03800}}].

\bibitem{CMSk2021jnp}
{\scshape CMS} collaboration, \emph{{Measurement of the top quark mass using
  events with a single reconstructed top quark in pp collisions at $ \sqrt{s} $
  = 13 TeV}}, \href{https://doi.org/10.1007/JHEP12(2021)161}{\emph{JHEP}
  {\bfseries 12} (2021) 161}
  [\href{https://arxiv.org/abs/2108.10407}{{\ttfamily 2108.10407}}].

\bibitem{CMS:2022emx}
{\scshape CMS} collaboration, \emph{{Measurement of the top quark pole mass
  using $ \textrm{t}\overline{\textrm{t}} $+jet events in the dilepton final
  state in proton-proton collisions at $ \sqrt{s} $ = 13 TeV}},
  \href{https://doi.org/10.1007/JHEP07(2023)077}{\emph{JHEP} {\bfseries 07}
  (2023) 077} [\href{https://arxiv.org/abs/2207.02270}{{\ttfamily
  2207.02270}}].

\bibitem{ATLAS:2017lox}
{\scshape ATLAS} collaboration, \emph{{Top-quark mass measurement in the
  all-hadronic $ t\overline{t} $ decay channel at $ \sqrt{s}=8 $ TeV with the
  ATLAS detector}}, \href{https://doi.org/10.1007/JHEP09(2017)118}{\emph{JHEP}
  {\bfseries 09} (2017) 118}
  [\href{https://arxiv.org/abs/1702.07546}{{\ttfamily 1702.07546}}].

\bibitem{CMS:2016iru}
{\scshape CMS} collaboration, \emph{{Measurement of the top quark mass using
  charged particles in pp collisions at $\sqrt s =$ 8 TeV}},
  \href{https://doi.org/10.1103/PhysRevD.93.092006}{\emph{Phys. Rev. D}
  {\bfseries 93} (2016) 092006}
  [\href{https://arxiv.org/abs/1603.06536}{{\ttfamily 1603.06536}}].

\bibitem{CMS:2016ixg}
{\scshape CMS} collaboration, \emph{{Measurement of the mass of the top quark
  in decays with a $J/\psi$ meson in pp collisions at 8 TeV}},
  \href{https://doi.org/10.1007/JHEP12(2016)123}{\emph{JHEP} {\bfseries 12}
  (2016) 123} [\href{https://arxiv.org/abs/1608.03560}{{\ttfamily
  1608.03560}}].

\bibitem{ATLAS:2014nxi}
{\scshape ATLAS} collaboration, \emph{{Measurement of the $t\bar{t}$ production
  cross-section using $e\mu $ events with b-tagged jets in pp collisions at
  $\sqrt{s}$ = 7 and 8 $\,\mathrm{TeV}$ with the ATLAS detector}},
  \href{https://doi.org/10.1140/epjc/s10052-016-4501-2}{\emph{Eur. Phys. J. C}
  {\bfseries 74} (2014) 3109}
  [\href{https://arxiv.org/abs/1406.5375}{{\ttfamily 1406.5375}}].

\bibitem{ATLAS:2022jbw}
{\scshape ATLAS} collaboration, \emph{{Measurement of the top-quark mass using
  a leptonic invariant mass in pp collisions at $ \sqrt{s} $ = 13 TeV with the
  ATLAS detector}}, \href{https://doi.org/10.1007/JHEP06(2023)019}{\emph{JHEP}
  {\bfseries 06} (2023) 019}
  [\href{https://arxiv.org/abs/2209.00583}{{\ttfamily 2209.00583}}].

\bibitem{CMS:2022kqg}
{\scshape CMS} collaboration, \emph{{Measurement of the differential $\hbox
  {t}\overline{\hbox {t}}$ production cross section as a function of the jet
  mass and extraction of the top quark mass in hadronic decays of boosted top
  quarks}}, \href{https://doi.org/10.1140/epjc/s10052-023-11587-8}{\emph{Eur.
  Phys. J. C} {\bfseries 83} (2023) 560}
  [\href{https://arxiv.org/abs/2211.01456}{{\ttfamily 2211.01456}}].

\bibitem{Holguin:2022epo}
J.~Holguin, I.~Moult, A.~Pathak and M.~Procura, \emph{{New paradigm for
  precision top physics: Weighing the top with energy correlators}},
  \href{https://doi.org/10.1103/PhysRevD.107.114002}{\emph{Phys. Rev. D}
  {\bfseries 107} (2023) 114002}
  [\href{https://arxiv.org/abs/2201.08393}{{\ttfamily 2201.08393}}].

\bibitem{Holguin:2023bjf}
J.~Holguin, I.~Moult, A.~Pathak, M.~Procura, R.~Sch\"ofbeck and D.~Schwarz,
  \emph{{Using the $W$ as a Standard Candle to Reach the Top: Calibrating
  Energy Correlator Based Top Mass Measurements}},
  \href{https://arxiv.org/abs/2311.02157}{{\ttfamily 2311.02157}}.

\bibitem{CMS:2024mlf}
{\scshape CMS} collaboration, \emph{{Measurement of energy correlators inside
  jets and determination of the strong coupling
  $\alpha_\mathrm{S}(m_\mathrm{Z})$}},
  \href{https://arxiv.org/abs/2402.13864}{{\ttfamily 2402.13864}}.

\bibitem{Basham:1978bw}
C.L.~Basham, L.S.~Brown, S.D.~Ellis and S.T.~Love, \emph{{Energy Correlations
  in electron - Positron Annihilation: Testing QCD}},
  \href{https://doi.org/10.1103/PhysRevLett.41.1585}{\emph{Phys. Rev. Lett.}
  {\bfseries 41} (1978) 1585}.

\bibitem{Dixon:2019uzg}
L.J.~Dixon, I.~Moult and H.X.~Zhu, \emph{{Collinear limit of the energy-energy
  correlator}}, \href{https://doi.org/10.1103/PhysRevD.100.014009}{\emph{Phys.
  Rev. D} {\bfseries 100} (2019) 014009}
  [\href{https://arxiv.org/abs/1905.01310}{{\ttfamily 1905.01310}}].

\bibitem{Lee:2022ige}
K.~Lee, B.~Me\c{c}aj and I.~Moult, \emph{{Conformal Colliders Meet the LHC}},
  \href{https://arxiv.org/abs/2205.03414}{{\ttfamily 2205.03414}}.

\bibitem{Chen:2020vvp}
H.~Chen, I.~Moult, X.~Zhang and H.X.~Zhu, \emph{{Rethinking jets with energy
  correlators: Tracks, resummation, and analytic continuation}},
  \href{https://doi.org/10.1103/PhysRevD.102.054012}{\emph{Phys. Rev. D}
  {\bfseries 102} (2020) 054012}
  [\href{https://arxiv.org/abs/2004.11381}{{\ttfamily 2004.11381}}].

\bibitem{Chen:2023zlx}
W.~Chen, J.~Gao, Y.~Li, Z.~Xu, X.~Zhang and H.X.~Zhu, \emph{{NNLL resummation
  for projected three-point energy correlator}},
  \href{https://doi.org/10.1007/JHEP05(2024)043}{\emph{JHEP} {\bfseries 05}
  (2024) 043} [\href{https://arxiv.org/abs/2307.07510}{{\ttfamily
  2307.07510}}].

\bibitem{CMS:2024yqd}
{\scshape CMS, ATLAS} collaboration, \emph{{Combination of measurements of the
  top quark mass from data collected by the ATLAS and CMS experiments at
  $\sqrt{s}=7$ and 8 TeV}},  \href{https://arxiv.org/abs/2402.08713}{{\ttfamily
  2402.08713}}.

\bibitem{Alwall:2014hca}
J.~Alwall, R.~Frederix, S.~Frixione, V.~Hirschi, F.~Maltoni, O.~Mattelaer
  et~al., \emph{{The automated computation of tree-level and next-to-leading
  order differential cross sections, and their matching to parton shower
  simulations}}, \href{https://doi.org/10.1007/JHEP07(2014)079}{\emph{JHEP}
  {\bfseries 07} (2014) 079} [\href{https://arxiv.org/abs/1405.0301}{{\ttfamily
  1405.0301}}].

\bibitem{Frederix:2018nkq}
R.~Frederix, S.~Frixione, V.~Hirschi, D.~Pagani, H.S.~Shao and M.~Zaro,
  \emph{{The automation of next-to-leading order electroweak calculations}},
  \href{https://doi.org/10.1007/JHEP11(2021)085}{\emph{JHEP} {\bfseries 07}
  (2018) 185} [\href{https://arxiv.org/abs/1804.10017}{{\ttfamily
  1804.10017}}].

\bibitem{Bierlich:2022pfr}
C.~Bierlich et~al., \emph{{A comprehensive guide to the physics and usage of
  PYTHIA 8.3}},
  \href{https://doi.org/10.21468/SciPostPhysCodeb.8}{\emph{SciPost Phys.
  Codeb.} {\bfseries 2022} (2022) 8}
  [\href{https://arxiv.org/abs/2203.11601}{{\ttfamily 2203.11601}}].

\bibitem{Cacciari:2008gp}
M.~Cacciari, G.P.~Salam and G.~Soyez, \emph{{The anti-$k_t$ jet clustering
  algorithm}}, \href{https://doi.org/10.1088/1126-6708/2008/04/063}{\emph{JHEP}
  {\bfseries 04} (2008) 063} [\href{https://arxiv.org/abs/0802.1189}{{\ttfamily
  0802.1189}}].

\bibitem{Cacciari:2005hq}
M.~Cacciari and G.P.~Salam, \emph{{Dispelling the $N^{3}$ myth for the $k_t$
  jet-finder}},
  \href{https://doi.org/10.1016/j.physletb.2006.08.037}{\emph{Phys. Lett. B}
  {\bfseries 641} (2006) 57}
  [\href{https://arxiv.org/abs/hep-ph/0512210}{{\ttfamily hep-ph/0512210}}].

\bibitem{Cacciari:2011ma}
M.~Cacciari, G.P.~Salam and G.~Soyez, \emph{{FastJet User Manual}},
  \href{https://doi.org/10.1140/epjc/s10052-012-1896-2}{\emph{Eur. Phys. J. C}
  {\bfseries 72} (2012) 1896}
  [\href{https://arxiv.org/abs/1111.6097}{{\ttfamily 1111.6097}}].

\bibitem{newville_2015_11813}
M.~Newville, T.~Stensitzki, D.B.~Allen and A.~Ingargiola, \emph{{LMFIT:
  Non-Linear Least-Square Minimization and Curve-Fitting for Python}},  Oct.,
  2015.
\newblock 10.5281/zenodo.11813.

\bibitem{delphes}
J.~de~Favereau, C.~Delaere, P.~Demin, A.~Giammanco, V.~Lemaître, A.~Mertens
  et~al., \emph{Delphes 3: a modular framework for fast simulation of a generic
  collider experiment},
  \href{https://doi.org/10.1007/jhep02(2014)057}{\emph{Journal of High Energy
  Physics} {\bfseries 2014} (2014) }.

\bibitem{CMS-DP-2020-019}
{\scshape CMS} collaboration, \emph{{Jet energy scale and resolution
  performance with 13 TeV data collected by CMS in 2016-2018}}, .

\end{thebibliography}\endgroup

\end{document}